# A conservative approach to leveraging external evidence for effective clinical trial design

Fabio Rigat, Oncology Biometrics, AstraZeneca Plc
fabio.rigat@astrazeneca.com

**Abstract**

Prior probabilities of clinical hypotheses are not systematically used for clinical trial design yet, due to a concern that poor priors may lead to poor decisions. To address this concern, a conservative approach to Bayesian trial design is illustrated here, requiring that the operational characteristics of the primary trial outcome are stronger than the prior. This approach is complementary to current Bayesian design methods, in that it insures against prior-data conflict by defining a sample size commensurate to a discrete design prior. This approach is ethical, in that it requires designs appropriate to achieving pre-specified levels of clinical equipoise imbalance. Practical examples are discussed, illustrating design of trials with binary or time to event endpoints. Moderate increases in phase II study sample size are shown to deliver strong levels of overall evidence for go/no-go clinical development decisions. Levels of negative evidence provided by group sequential confirmatory designs are found negligible, highlighting the importance of complementing efficacy boundaries with non-binding futility criteria.

# Introduction

Clinical trial design focuses on providing novel evidence about clinical hypotheses [1-3]. To this end, historical and real-world data provide a basis to define what is known about the primary endpoint at design stage. This background evidence also complements data from multiple clinical trials when assessing the overall safety and efficacy of treatments using Bayesian meta-analysis [4-9]. Likewise, Bayesian trial design combines prior data with the likelihood of the yet unobserved trial outcome to inform design decisions based the totality of the data that will be available after the trial primary analysis [10-14]. Hence, a specific focus of Bayesian design is the control of the overall probability of false positive findings when an optimistic prior is combined with negative trial outcomes. A common approach to false positive error control here is to down-weight priors used as a basis for trial design when these are found to conflict with the trial data. This perspective is reinforced here, by requiring that the operating characteristics of the primary trial outcome dominate the design prior. Hence, here stronger priors require commensurately greater sample sizes to ensure that overall evidence about the design hypotheses is strong, regardless of whether the trial outcome is positive or negative. The practical value of this conservative approach is presented in examples using binary and time-to-event endpoints commonly accepted for oncology studies. Moderate increases in sample size are shown to deliver strong levels of overall evidence informing robust "go/no-go" decisions based on the totality of phase I and phase II trial data. Levels of negative evidence provided by group sequential confirmatory designs are found negligible, highlighting the importance of complementing efficacy boundaries with non-binding futility criteria.

## 1. Bayesian inference quantifies evidence about clinical trial hypotheses

Clinical trials are designed to ensure strong control of the probability of false positive outcomes (low type 1 error or, equivalently, high specificity) [15]. Subject to this specificity requirement, study sample sizes are calculated to provide high true positive probability (high power or sensitivity) under specific design scenarios. However, it is well known that these operating characteristics alone cannot bear information about which design hypothesis is best supported by the observed trial outcome, because they are calculated conditionally on the null or on the alternative hypothesis being true. The truth probability of clinical hypotheses given trial outcomes can be formally estimated using Bayesian inference [16-20], which is applied to design and analysis of paediatric trials, rare diseases, dose escalation and dose-response trials, adaptive trials [21], medical diagnostics [22] and in clinical practice [23-24].

To illustrate Bayesian design in essence, here $H_1$ and $H_0$ represent respectively the motivating trial hypothesis and the associated null. Typically, $H_1$ denotes superior clinical benefit of an investigational treatment over standard of care and $H_0$ represents the lack of such additional benefit. The symbols "+" and "-" here indicate respectively a positive and a negative outcome of the primary endpoint analysis with pre-specified specificity $p(-|H_0)$ and sensitivity $p(+|H_1)$. The outcome of a confirmatory study is traditionally defined positive when its p-value is smaller than a pre-specified threshold, no greater than 5%, defining the false positive probability of the observed treatment effect. When Bayesian inference is used, the study outcome is positive when the posterior probability of the treatment effect exceeding a pre-specified clinically relevant threshold is much greater than the prior, with typical thresholds exceeding 95%. The design approach described here may also be applied to settings allowing for a more nuanced classification of the study outcome (e.g. [25]).

Here $P(H_0)$ and $P(H_1)$ represent the prior probabilities of the design hypotheses about the trial primary endpoint. Their ratio

$$r_{01} = \frac{P(H_0)}{P(H_1)}. \qquad (1)$$

defines the pre-study odds of the design hypotheses. The weakest design prior $r_{01} = 1$ denotes equi-probable clinical hypotheses. The values $r_{01} > 1$ and $r_{01} < 1$ denote respectively stronger pre-study evidence in favour of $H_0$ or of $H_1$. Note here that the prior $r_{01}$ is used for study design only, and not for analysis of the study data nor for establishing whether the study outcome is positive or negative. Also, in practice design prior probabilities are estimated from published clinical trial results, real-world data and through prior elicitation [26]. However, in absence of standard processes for derivation of these priors, the notation used here omits dependence from specific historical data or prior hyperparameters.

The overall evidence provided by the design prior together with the primary trial outcome is quantified by the post-study odds of the design hypotheses

$$r_{01}(-) = \frac{P(H_0|-)}{P(H_1|-)} = r_{01} \times \frac{p(-|H_0)}{1-p(+|H_1)}, \qquad (2)$$

$$r_{10}(+) = \frac{P(H_1|+)}{P(H_0|+)} = r_{10} \times \frac{p(+|H_1)}{1-p(-|H_0)}, \qquad (3)$$

where $r_{10} = 1/r_{01} = P(H_1)/P(H_0)$. The posterior odds on left-hand side of (2) and (3) are herein referred to as Bayesian characteristics (BACs) of the trial design. The negative and positive likelihood ratios multiplying the pre-study odds on the right-hand-side of (2) and (3) respectively measure the change in pre-study odds upon observing a negative or a positive study outcome with pre-specified specificity and sensitivity. When these likelihood ratios are calculated by averaging over parameters indexing the endpoint distribution, they are referred to as Bayes factors [27]. Analogously to the pre-study odds, greater values of (2) and (3) provide stronger support respectively for $H_0$ and for $H_1$. Also, differences in scale between design prior and operating characteristics of the trial outcome are irrelevant to calculating or interpreting the post-study odds, which are defined as products of ratios and hence are unitless.

## 1.1 Strong BACs design prevents evidence regression

Trial design prevents *evidence regression* when the overall evidence about the trial hypotheses upon observing the trial outcome is ensured to be stronger than the background evidence available a priori, regardless of whether the trial outcome is positive or negative. BACs design prevents evidence regression at pre-specified overall evidence thresholds $(\tau_N, \tau_P)$ when

$$\begin{cases} r_{01}(-) \geq \tau_N > r_{01}, \\ r_{10}(+) \geq \tau_P > r_{10}. \end{cases} \qquad (4)$$

Condition (4) is in fact weak, in that it only requires post-study odds to increase beyond the pre-study odds, which in practice may be close to equi-probability. By also requiring that

$$\begin{cases} \tau_N > \max(1, r_{01}), \\ \tau_P > \max(1, r_{10}), \end{cases} \qquad (5)$$

the post-study odds cannot be in favour of $H_1$ upon observing a negative trial outcome, no matter how optimistic the design prior. Likewise, under (5) the post-study odds cannot be in favour of $H_0$ when the trial is positive, no matter how pessimistic the design prior. Design fulfilling (5) is henceforth referred to as strong BACs design, because evidence regression is effectively prevented and the trial outcome is conclusive at levels $(\tau_N, \tau_P)$, as depicted in Figure 1. The grey zone here marks the range representing insufficient evidence about the design hypotheses. The red and green zones in Figure 1 mark respectively the range of post-study odds associated to strong negative or strong positive evidence at levels $(\tau_N, \tau_P)$.

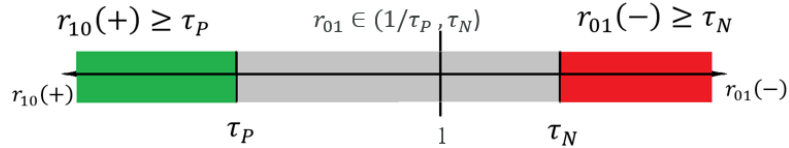

*Figure 1: weak evidence about the design hypotheses (grey zone), reflecting pre-study odds close to 1. Strong post-study evidence is associated respectively with large negative (red zone) or large positive (green zone) post-study odds.*

Under strong BACs design the specificity and sensitivity of the study outcome are calibrated to ensuring that the post study odds of the design hypotheses cannot be dominated by the design prior $r_{01}$. Hence, (5) prevents by design the possibility to pre-determine overall evidence in favor of against clinical hypotheses using a strong prior in conjunction with weak operational characteristics. For instance, when the prior favours the null ($r_{01} > 1$), (3) and (5) require a stronger positive likelihood ratio to disprove $H_0$ relative to $H_1$ compared to equi-probability of the trial hypotheses. Likewise, (2) and (5) require a stronger negative likelihood ratio to disprove a promising alternative ($r_{10} > 1$) compared to using a weaker prior. To this end, since specificity $p(-|H_0)$ in confirmatory trials is typically high, strong BACs design using informative priors will require greater power compared to equi-probable hypotheses. Hence, strong BACs design provides a mechanism to discourage re-testing of established clinical facts ($r_{01} \gg 1$ or $r_{01} \ll 1$), as this would require prohibitively large sample sizes.

Of note, strong BACs design also protects against sources of bias that cannot be avoided when the operating characteristics of the trial outcome are not commensurate to the background evidence, including base rate fallacy, availability bias and confirmation bias [28].

**1.2 Strong BACs design is ethical**

Equipoise has been proposed as an ethical basis for enrolment of participants in clinical trials [29]. Clinical equipoise was defined as a state of insufficient evidence about the relative benefit of competing treatments resulting in "professional disagreement among expert clinicians about the preferred treatment" [30]. Clinical equipoise is thus the quantitative basis of design approaches regarding the primary objective of a trial as the accrual of evidence adequate to shift collective medical judgement and health policy decisions out of unacceptable uncertainty [31-35]. When equipoise is adopted as a basis for enrolment of patients in a clinical trial, the pre-study odds $r_{01}$ quantify the degree of clinical equipoise present at design stage as estimable from background data. Strong BACs design here ensures that the operating characteristics of the primary trial outcome are adequate to demonstrating clinical equipoise imbalance greater than the pre-specified evidence thresholds $(\tau_N, \tau_P)$. Hence, (2)-(5) enable achieving the goal of disturbing an initial state of equipoise through an actionable trial design methodology.

**1.3 BACs evidence thresholds for clinical trial design**

Table 1 compares the operational characteristics of trial outcomes to the associated BACs under illustrative scenarios relevant to early and late development. Effect sizes and sample sizes associated to different sensitivity (power) values are not reported in Table 1, although specific assumptions are needed to calculate these operational characteristics in practice. Results in Table 1 can also be generalised to cases when a prior probability distribution for the pre-study odds $r_{01}$ is used [36].

First Table 1 shows that a trial outcome with 50% probability of being positive or negative under either $H_0$ or $H_1$, thus being both underpowered and unable to control false positive errors, cannot provide any new evidence on the BACs scale. This is a basic requirement for any evidence scale to be useful in practice. Of note, it follows from (2)-(3) that $P(H_0|-) \geq P(H_0)$ and $P(H_1|+) \geq P(H_1)$ when specificity and sensitivity of the study outcome are greater than 50%. Equivalently, BACs provide meaningful probabilistic support [37] to any design that is not completely uninformative.

Second, Table 1 shows the BACs values at 95% specificity and 80% sensitivity for equi-probable design hypotheses. A negative trial outcome here provides overall evidence in favour of the null 4.75 times stronger than for the alternative, and a positive outcome provides overall evidence for the alternative 16 times stronger than for the null. Positive evidence is over three times stronger than the negative evidence level because the probability of false positive outcomes (5%) is smaller than the probability of false negatives (20%). As equi-probability $r_{01} = 1$ is the weakest design prior and late phase clinical studies are commonly designed with 95% specificity and at least 80% sensitivity, the BACs thresholds $r_{01}(-) = 4.75$ and $r_{10}(+) = 16$ quantify the lower bounds of positive and negative overall evidence levels consistent with current confirmatory trial design practices.

Table 1 next shows that, at 90% specificity and 90% sensitivity and for equi-probable design hypotheses, evidence in favour of the null/alternative hypothesis ensuing from a negative/positive outcome is symmetric, taking value 9 either way. BACs symmetry holds whenever the design hypotheses are equally a priori likely, and specificity and sensitivity of the trial outcome have equal values.

Bounds for equipoise imbalance acceptable as a basis for enrolling patients in clinical trials were reported within the range 60%-70% [38], that is $r_{01} \in (0.38, 0.41)$ or, equivalently, $r_{10} \in (2.44, 2.63)$. Table 1 shows the BACs calculated for design priors within these bounds, demonstrating the extent to which stronger operational characteristics are required for strong BACs design when using design priors other than equi-probability. At 95% specificity and 80% sensitivity, using the optimistic prior $r_{01} = 1/2$ the overall evidence combining the design prior with a negative trial outcome is halved ($r_{01}(-) = (1/2) \times 0.95/0.2 = 2.38$) and that from a positive outcome is doubled ($r_{10}(+) = 2 \times 0.8/0.05 = 32$) compared to equi-probable hypotheses. Here an increase in power to 90% barely restores strong overall negative evidence $r_{01}(-) = 4.75$ preserving strong positive evidence ($r_{10}(+) = 36$). The sample size needed to increase power from 80% to 90% will depend in practice on the distributional properties of the primary endpoint, interim analyses, enrolment, and dropout rates. For instance, a trial with time-to-event primary endpoint and two interim analyses planned using group sequential design [39] requires roughly a 30% sample size increase to raise power from 80% to 90%. Using the pessimistic pre-study odds $r_{01} = 2$ again at 95% specificity and 80% sensitivity, evidence from a positive outcome is halved ($r_{10}(+) = 8$) and that from a negative trial outcome is doubled ($r_{01}(-) = 9.5$) compared to equi-probability. However, here strong positive evidence ($r_{10}(+) \geq 16$) cannot be restored by increasing power alone. For instance, when the pessimistic design prior $r_{01} = 2$ is used and specificity is 95%, the upper bound of the overall positive evidence is $r_{10}(+) < 10$. Hence, under $r_{01} = 2$ strong overall positive and negative evidence can only be ensured by design by increasing both specificity and sensitivity of the trial outcome, up to 97.5% and 90% respectively ($r_{01}(-) = 19.5, r_{10}(+) = 18$). To put in perspective the cost entailed by use of this pessimistic design prior, a time to event trial with interim analyses using group sequential design would require roughly a 50% sample size increase to raise power from 80% to 90% and decrease type 1 error from 5% to 2.5%. Hence, Table 1 suggests that in practice strong overall evidence under optimistic design priors can be achieved when a moderate sample size increase is acceptable. Use of pessimistic design priors may require both smaller type 1 error levels and greater power, making these priors impractical compared to equi-probable design hypotheses. However, use of pessimistic priors beyond phase I is unlikely in practice, as this typically requires positive pre-study evidence.

| Pre-study odds $r_{01}$ | Operating characteristics | | BACs | |
|---|---|---|---|---|
| | Specificity | Sensitivity | $r_{01}(-)$ | $r_{10}(+)$ |
| any real value | 50% | 50% | $r_{01}$ | $1/r_{01}$ |
| 1 | 95% | 90% | 9.5 | 18 |
| 1 | 95% | 80% | 4.75 | 16 |
| 1 | 90% | 90% | 9 | 9 |
| 1 | 80% | 80% | 4 | 4 |
| 1/2 | 95% | 90% | 4.75 | 36 |
| 1/2 | 95% | 80% | 2.38 | 32 |
| 1/2 | 90% | 90% | 4.5 | 18 |
| 1/2 | 80% | 80% | 2 | 8 |
| 2 | 95% | 90% | 19 | 9 |
| 2 | 95% | 80% | 9.5 | 8 |
| 2 | 90% | 90% | 18 | 4.5 |
| 2 | 80% | 80% | 8 | 2 |

Table 1: BACs (right) of design at selected values of the pre study odds $r_{01}$, compared to the associated operating characteristics

## 2. Practical BACs design

This Section considers applications of strong BACs design to two clinical development settings. First, BACs are applied to phase II study design. Here investment in designs providing strong positive or negative evidence, compatibly with safeguarding patient safety, rewards sponsors by providing robust go/no-go late development decisions based on the totality of the efficacy data observed in phase I and in phase II. Second, BACs are applied in a confirmatory design setting to quantify the strength of positive and negative evidence provided by standard group sequential design with time to event endpoints.

### 2.1 Simon 2-stage phase II design leveraging phase I efficacy data

Clinical development is considered here within a setting where patients do not have efficacious treatment options, with historical data showing 10% objective response rate (ORR) under standard of care. Phase 1 ORR data shows 2 responders among 9 subjects treated (ORR = 22%) with an investigational treatment. Figure 2 shows the *Beta(3,8)* ORR posterior probability density, estimated by combining a uniform ORR prior on 0-100% with the Binomial likelihood of the phase I data. The 95% equal-tails credible interval 7%-56% encompasses a broad uncertainty about ORR typical of phase I studies designed with safety primary endpoints.

Since ORR is directly attributable to drug effect, single-arm trials are a viable design option to assess ORR in populations where no efficacious therapy exists [40]. To this end, Simon 2-stage (S2S, [41]) is used for phase II design with ORR primary endpoint. Given pre-specified unacceptable (URR) and target (TRR) response rates, S2S tests $H_0$: ORR $\leq$ URR versus $H_1$: ORR $\geq$ TRR. The URR is set at the historical 10% rate, marked in red in Figure 2. Here the ORR mean, median and mode are respectively 27%, 26% and 22%, reflecting a right-skewed distribution. The posterior mode, marked in black in Figure 2, is selected to defining the TRR = 22% level for the phase II study design. This is the most conservative TRR option among the three population summaries, mitigating the risk of overestimating the end-stage population ORR due to lack of randomisation in phase I. The pre-study odds for the phase II ORR primary endpoint are calculated as the ratio of the phase I posterior probability densities $r_{01} = f(URR)/f(TRR) \approx 0.56$. Table 2 compares the operating characteristics of optimal S2S designs to the BACs at the pre-study odds $r_{01} = 0.56$ and $r_{01} = 1$. The top row shows that a negative S2S outcome at 95% specificity, 80% power and maximum sample size $N_f = 66$ is insufficient to overwhelm the optimistic pre-study odds at the BACs thresholds $\tau_N = 4.75, \tau_P = 16$. The bottom row in Table 2 shows that the S2S design at 90% power and $N_f = 98$ provides strong overall positive and negative evidence when combined to the pre-study odds $r_{01} = 0.56$. Hence, the BACs here demonstrate that an increase of (98-66)/66 ≈ 49% maximum S2S sample size provides robust evidence for late development planning based on the totality of the ORR data observed both in phase I and in the phase II study.

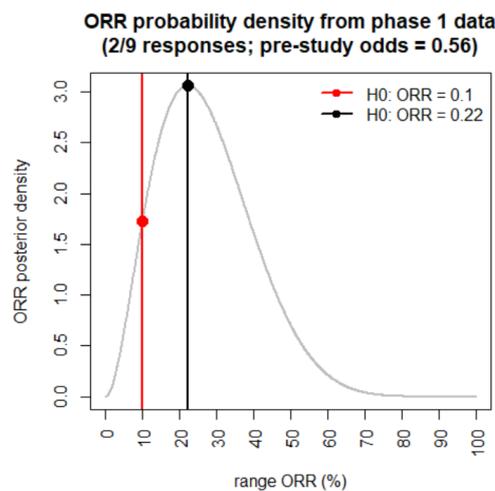

*Figure 2: posterior probability density of the ORR estimated by combining a uniform prior with 2 responses among 9 participants of an illustrative single arm phase I study.*

| Operating Characteristics | | Optimal Simon 2 stage design | | | | | | BACs | | | |
|---|---|---|---|---|---|---|---|---|---|---|---|
| specificity | sensitivity | $N_1$ | $R_1$ | $N_F$ | $R_F$ | EN | PET | $r_{01} = 0.56$ | | $r_{01} = 1$ | |
| | | | | | | | | $r_{01}(-)$ | $r_{10}(+)$ | $r_{01}(-)$ | $r_{10}(+)$ |
| 95% | 80% | 21 | 2 | 66 | 10 | 37 | 65% | 2.7 | 21.6 | 4.75 | 16 |
| | 85% | 29 | 3 | 73 | 11 | 44 | 67% | 3.6 | 30.3 | 6.3 | 17 |
| | 90% | 36 | 4 | 98 | 14 | 54 | 71% | 5.3 | 32.1 | 9.5 | 18 |

*Table 2: operating characteristics and BACs of Simon 2-stage (S2S) optimal designs testing $H_0$: ORR $\leq$ 10% versus $H_1$: ORR $\geq$ 22% at the phase I pre-study odds $r_{01} = 0.56$ and under equi-probable design hypotheses. Strong overall evidence is achieved by the S2S design carrying 95% specificity and 90% sensitivity.*

## 2.2 Single arm and single analysis phase II trial design

Unlike for the one-sided hypotheses used in Section 2.1, the single arm phase II study design here tests the sharp null $H_0: ORR = 10\%$ against the alternative $H_1: ORR = 22\%$. A single efficacy analysis is pre-specified for the ORR primary endpoint and overall evidence is assessed again based on the pre-study odds $r_{01} = 0.56$ estimated from the phase I response data. The phase II outcome is defined positive when the Binomial p-value under $H_0$ at the observed ORR is smaller than 5%, ensuring control of the false positive error rate. Table 3 shows that the sample size N = 70 ensures 95% specificity and approximately 80% sensitivity of this outcome. Hence, N = 70 is a sample size associated to acceptable operating characteristics for this single arm design. However, these operating characteristics do not contain information about the phase I data beyond the point summary of the ORR distribution defining the design hypotheses. When the pre-study odds in favour of $H_1$ are considered, Table 3 shows that an outcome based on N = 70 participants is not sufficiently powered to providing strong overall negative phase II evidence at the BACs thresholds $\tau_N = 4.75, \tau_P = 16$. Table 3 shows that increasing sample size by (95-70)/70 ≈ 36% is needed to providing robust positive and negative evidence for late development go/no-go decisions using both phase I and phase II ORR data.

| N | Operating characteristics | | BACs at $r_{01} = 0.56$ | |
|---|---|---|---|---|
| | Specificity | sensitivity | $r_{01}(-)$ | $r_{10}(+)$ |
| 70 | 95% | 79% | 2.6 | 28.1 |
| 75 | | 82% | 2.9 | 29.1 |
| 80 | | 84% | 3.4 | 30 |
| 85 | | 86% | 3.9 | 30.8 |
| 90 | | 88% | 4.6 | 31.4 |
| 95 | | 90% | 5.4 | 32 |

*Table 3: sample size (N), operating characteristics and BACs for a single arm study testing the hypotheses $H_0: ORR = 10\%$ vs $H_1: ORR = 22\%$ when the pre-study odds are $r_{01} = 0.56$, reflecting prior evidence in favour of $H_1$.*

## 2.3 Randomized phase II study

To prevent potential biases ensuing from lack of randomisation in single arm trials, a parallel arms phase II randomised study design is considered here testing $H_0$: ORR = 10% in both arms versus $H_1$: ORR = 22% in the investigational arm. A 1:1 randomization ratio is used, and the study outcome is defined positive is the p-value of the Chi-square test statistic comparing the ORR between the two study arms is smaller than 5%. Figure 3 depicts the overall positive $r_{10}(+)$ and negative $r_{01}(-)$ evidence levels for this design plotted against the associated power at 95% specificity. Blue dots mark values associated to the weakest design prior $r_{01} = 1$, and red dots mark values calculated using the pre-study odds $r_{01} = 0.56$ estimated from the phase I ORR distribution in Figure 2. At the reference BACs thresholds $\tau_N = 4.75, \tau_P = 16$, marked as thick horizontal in Figure 3, strong operational characteristics and BACs are achieved respectively at N = 300 and N = 370 total sample size at $r_{01} = 1$ and $r_{01} = 0.56$. Hence, the BACs here demonstrate that an increase in study sample size of (370-300)/300 ≈ 23% is sufficient to protect late phase go/no-go decisions against inconsistencies between the optimistic single arm phase I data and the outcome of a randomised proof of concept study.

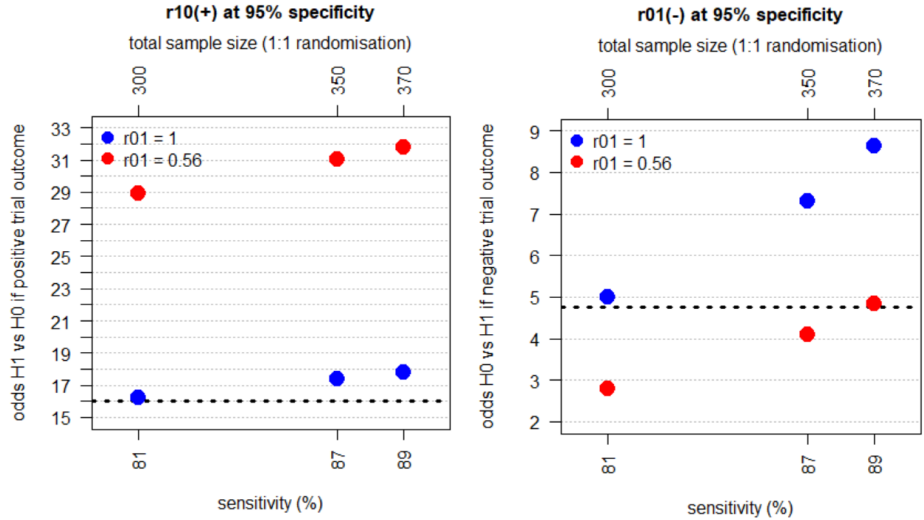

*Figure 3: overall negative (left) and positive (right) evidence by trial sample size and associated sensitivity of 1:1 randomised phase II study design with ORR primary endpoint at 95% specificity. Using the optimistic pre-study odds $r_{10} = 0.56$, strong operational characteristics and BACs are achieved at a total sample size of 370 subjects.*

## 2.4 Group sequential design with time-to-event endpoint

Group sequential (GS) designs enable pre-planning of interim analyses under strong control of the family-wise false positive error rate. BACs are used here to evaluating the strength of positive and negative evidence provided by GS designs under the O'Brien and Fleming and the Pocock alpha spending boundaries. Perfect clinical equipoise ($r_{01} = 1$) is assumed, reflecting weak pre-study evidence about time-to-event endpoints in the target population. Efficacy is measured by the hazard ratio (HR) between investigational and control arms, with $H_0: HR = 1$ and $H_1: HR = 0.67$, so that the trial is designed to detect a 33% reduction in event rate among subjects at risk relative to the controls. The median time to event in the control arm is 10 months. When events times are exponentially distributed, the median time to event in the investigational arm is expected to be 15 months under $H_1$, representing a 50% increase compared to control. Expected duration of enrolment is 24 months with 6 months ramp-up to the steady state enrolment rate, and minimum follow-up is 12 months. Event times are analysed using the log-rank test, and one interim analysis is planned at 50% aggregate maturity.

Table 4 shows the GS design sample size and operating characteristics, calculated using the R package *gsDesign* [42]. For comparison, the bottom line in Table 4 reports the sample size of the fixed design with no interim analyses at 95% specificity and 90% sensitivity. Crossing of the GS HR boundary provides stronger positive evidence on the BACS scale in favour of the alternative both at interim and at final analysis compared to the fixed design ($r_{10}(+) > 18$ versus $r_{10}(+) = 18$). Failure to cross the HR interim boundary provides weak evidence in favour of the null ($r_{01}(-) \leq 2.85$), due to the low power associated to the interim GS boundaries. This weakness highlights the practical importance of complementing the GS efficacy boundaries with non-binding futility boundaries, consistently with regulatory guidelines [21]. The HR futility boundary at interim is HR = 0.955 using O'Brien and Fleming and HR = 0.852 using Pocock alpha spending. The associated futility probabilities are respectively 60.5% and 84.5% under $H_0$ and 2% and 6% under $H_1$ using O'Brien and Fleming and the Pocock boundaries. On the BACs scale, evidence in favor of $H_0$ is increased by 30-fold and by 14-fold when the HR crosses respectively the O'Brien and Fleming and the Pocock interim futility boundaries. These BACS complement the operational characteristics of the futility analysis, showing that O'Brien and Fleming boundaries provide strong evidence to assess whether enrolment should be stopped to prevent further patients being assigned to treatments unlikely to provide benefit beyond standard of care.

| Spending function | Analysis Stage | N. events | N. subjects enrolled | Operating characteristics | | HR boundary | BACs | |
|---|---|---|---|---|---|---|---|---|
| | | | | specificity (%) | sensitivity (%) | | $r_{01}$ (-) | $r_{10}$ (+) |
| O'Brien-Fleming | IA | 134 | 366 | 99.7 | 26 | 0.599 | 1.35 | 86.8 |
| | FA | 268 | 416 | 95.1 | 90 | 0.786 | 9.51 | 18.5 |
| Pocock | IA | 165 | 450 | 96.9 | 66 | 0.714 | 2.85 | 21.3 |
| | FA | 329 | 510 | 95.4 | 90 | 0.784 | 9.54 | 19.6 |
| Fixed design | | 262 | 406 | 95 | 90 | 0.783 | 9.5 | 18 |

*Table 4: operational characteristics of group sequential designs using the O'Brien and Fleming and the Pocock boundaries and associated BACs under equi-probable hypotheses.*

## 3. Discussion

Falsifiability is the accepted criterion demarcating scientific models from dogma [43-44]. Clinical trial design priors are falsifiable using published Bayesian models allowing for prior updating based on its discrepancy with the study outcome. When these methods are used, the study sample size needs to be sufficient to ensuring that the study outcome is conclusive even when prior-data conflict is observed. Strong BACs design provides a conservative solution to this issue, motivating the design of studies endowed with sufficient evidential strength to overcome prior-data conflict. Current limitations to effective implementation of BACs design include limited access to background data and lack of standard methods to collect, analyse and synthesize multiple sources of background evidence for trial design. These barriers are becoming less challenging [45], thanks to more systematic publication of clinical trials reporting positive and negative outcomes, to the increasing availability of real-world data and to the development of guidelines enabling their use in clinical development. When prior probabilities of clinical hypotheses can be derived from these background data, use of strong BACs design ensures that use of design priors cannot lead to weaker decisions compared to designs based on conditional error probabilities only. From this perspective, no measure of statistical scepticism can lead to logically objecting to commensurate study designs which operating characteristics are calibrated to dominate the priors. However, the price paid is high, as strong insurance against prior-data conflict leads to greater sample sizes compared to traditional designs. This increase in sample size may not be achievable in some clinical development settings, including rare diseases and in vulnerable populations. In this context, assessment of overall evidence thresholds associated to feasible sample sizes may be of practical value. In general, more flexible methodology is needed encompassing consideration of minimum evidence levels needed at different development stages and degree of insurance against possibly unreliable priors.

### Research ethics and consent

No ethics committee approval and no informed consent were required because no clinical trial data were used throughout.

### Competing interests

None.

## Acknowledgements

I am grateful to many Colleagues who reviewed earlier versions of this work, including Peter De Porre, Stephen Weng, Telba Irony, Vladimir Dragalin, Tobias Mielke, Dwaine Banton, Filip De Ridder, Huiming Liao, Rafael Sauter, Maoji Li, Fowzia Ibrahim and Binbing Yu. Insightful comments were also provided by Professor David Spiegelhalter (Cambridge) and Professor Jim Smith (Warwick).

## Funding

This work was completed while the Author was full time employee of AstraZeneca PLC Research and Development.

## Data sharing statement

Not applicable